\documentclass[11pt,a4paper]{article}
\usepackage{jheppub}
\usepackage{graphicx,multirow}

\newcommand{\comment}[1]{}

\newcommand{\beq}{\begin{equation}}
\newcommand{\eeq}{\end{equation}}

\newcommand{\pol}{{\cal P}}

\begin{document}

\title{Kinked tracks from $\Sigma^+$ baryons as a probe of light quark polarizations}

\affiliation{Department of Physics, Ben-Gurion University, Beer-Sheva 8410501, Israel}

\author{Yevgeny Kats}
\emailAdd{katsye@bgu.ac.il}

\abstract{Jets initiated by energetic quarks carry in an obvious way approximate information about the original quark energy and momentum. In a less obvious way, they also carry information about the quark polarization. Previous works have shown how the polarization information can be extracted by ATLAS and CMS using events in which the quark hadronizes to an energetic baryon --- $\Lambda_b$ for $b$ quarks, $\Lambda_c$ for $c$ quarks and $\Lambda$ for $s$ quarks. In this paper we extend these proposals to $\Sigma^+$ baryons, which can provide sensitivity to polarizations of $s$ and $u$ quarks. We analyze the various aspects of how the kinked track signature of the $\Sigma^+ \to p \pi^0$ decay can be used for this purpose. We evaluate the feasibility of such measurements in $t\bar t$ samples.}

\maketitle

\section{Introduction and motivation}

When a baryon is one of the leading hadrons in a jet, it partly preserves the polarization of the quark that produced the jet. From the theory side, this can be argued rigorously for heavy quarks, $m_q \gg \Lambda_{\rm QCD}$~\cite{Mannel:1991bs,Ball:1992fw,Falk:1993rf,Galanti:2015pqa}, and it is expected to hold also for light quarks, as we will review. From the experiment side, the LEP experiments have already observed the polarizations transferred to the $\Lambda_b$~\cite{Buskulic:1995mf,Abbiendi:1998uz,Abreu:1999gf} and $\Lambda$~\cite{Buskulic:1996vb,ALEPH:1997an,Ackerstaff:1997nh} baryons from the longitudinally polarized quarks produced in $Z$ decays. In the case of $\Lambda$ baryons produced from strange quarks, the polarization transfer was found to be significant when the baryon carried a sizable fraction ($\gtrsim 30\%$) of the quark momentum.

There exist at least two motivations for measuring baryon polarizations in samples of polarized quarks. First, these measurements contribute to the theoretical understanding, or at least the phenomenological description, of the spin structure of baryons and the fragmentation process. This is important both from the purely theoretical perspective and for tuning Monte Carlo generators. Second, once techniques for measuring the baryon polarizations, and relations between the quark and baryon polarizations, are established, it becomes possible to measure polarizations of quarks in new processes.

Measurements of the transverse $\Lambda_b$ polarization from QCD production were performed by CMS~\cite{Sirunyan:2018bfd} and LHCb~\cite{LHCb:2020iux}, and similarly for the $\Lambda$ by ATLAS~\cite{ATLAS:2014ona}. However, extracting information from such measurements is difficult because the transverse polarization of quarks in QCD samples is an NLO effect, which is sizable only for soft quarks and depends strongly on the kinematics of the event. Additionally, different from the case of longitudinal polarization, parity invariance allows transverse polarization to be generated in the hadronization process even for initially unpolarized quarks~\cite{Mulders:1995dh,Anselmino:2000vs}, further complicating the interpretation.

Longitudinally polarized quarks produced in electroweak processes can therefore be a more convenient target. In this regard, it has been argued~\cite{Galanti:2015pqa,Kats:2015cna} that meaningful measurements of the longitudinal $\Lambda_b$, $\Lambda_c$ and $\Lambda$ polarizations due to the $b$, $c$ and $s$ quarks from top or $W$ decays in $t\bar t$ samples are possible in the Run~2 datasets of ATLAS and CMS. The statistics of such polarized quark samples is already comparable to that of the hadronic $Z$ decays at LEP. The initial quark polarization in the $t\bar t$ samples (before the small one-loop QCD effects~\cite{Korner:1993dy}) is close to $-1$, while at LEP it was about $-0.94$ for the down-type quarks and $-0.67$ for the up-type ones. For the $\Lambda_b$ it will be the first time to obtain precision of order $10\%$, and for the $\Lambda_c$ the first time to measure longitudinal polarization of energetic charm quarks~\cite{Galanti:2015pqa}. For the $\Lambda$, it will be the first time to experimentally disentangle the contributions of the different quark flavors~\cite{Kats:2015cna}. Another promising sample is $\Lambda_c$ baryons from $W$+$c$ production~\cite{Kats:2015zth}. In the current work we extend these studies to the interesting case of the $\Sigma^+$ baryon.

There are several reasons to consider the $\Sigma^+$ (whose valence quark content is $uus$) despite its similarity to the $\Lambda$ ($uds$). First, the probability for a strange quark to hadronize to a $\Sigma^+$ is lower than to a $\Lambda$ by only a factor of about~2 (in the relevant range of momentum fractions, which we will take to be $z > 0.3$)~\cite{Bourrely:2003wi}. Thus, if a spin-sensitive decay of the $\Sigma^+$ can be reconstructed with a sufficiently high efficiency and low background, it will be possible to reduce the statistical uncertainty of the strange-quark polarization measurement relative to what can be obtained~\cite{Kats:2015cna} by using the $\Lambda$ alone. The reconstruction efficiency of the $\Lambda$ decays in ATLAS and CMS in the relevant range of $p_T$ is somewhat low, ${\cal O}(10\%)$~\cite{ATLAS:2014ona}, due to their very large displacement. The $\Sigma^+$ decays, which are less displaced by a factor of~3.3, might therefore be competitive. We consider the decay
\beq
\Sigma^+ \to p \pi^0 \,,
\label{decay}
\eeq
whose branching fraction is about $52\%$ (cf.\ $64\%$ for $\Lambda \to p\pi^-$) and spin analyzing power\footnote{The spin analyzing power, which is also called the decay asymmetry parameter, describes the sensitivity of the angular distribution of the decay to the hadron polarization; see eq.~\eqref{spin-analyzing} below.} $\alpha \approx -0.98$ (cf.\ $0.73$ for $\Lambda\to p\pi^-$)~\cite{PDG}. The $\pi^0$ has the disadvantage of not creating tracks (but only calorimeter deposits). On the other hand, the possibility to measure the $\Sigma^+$ track directly presents an opportunity, as we will discuss. The other important decay mode, $\Sigma^+ \to n\pi^+$, is not useful since it has a much lower spin analyzing power, of about $0.07$~\cite{PDG}, and at the same time is not obviously much easier to reconstruct. For the same reason we do not consider the $\Sigma^-$, which decays almost $100\%$ as $\Sigma^- \to n\pi^-$, with a spin analyzing power of about $-0.07$~\cite{PDG}. The $\Sigma^0$ baryon decays almost entirely as $\Sigma^0 \to \Lambda\gamma$. Since the resulting prompt photon is very soft, it is difficult to isolate $\Sigma^0$ events from the inclusive $\Lambda$ events, therefore we do not consider the $\Sigma^0$ either.

Even in the case that the statistics of the $\Sigma^+$ samples from strange quarks ends up somewhat lower than for the analogous $\Lambda$ samples, an even more interesting reason for measuring the $\Sigma^+$ is that it can provide qualitatively different information, especially in the context of the $u$ quark polarization. In the na\"{i}ve quark model, the spin of the $\Lambda$ is carried entirely by the $s$ quark, while the $u$ and $d$ form a spin-0 diquark. This suggests that $\Lambda$ baryons produced in the hadronization of polarized $u$ or $d$ quarks do not retain much of the polarization. Differently, in the $\Sigma^+$, still in the na\"{i}ve quark model, the two $u$ quarks form a spin-1 diquark, so one can expect $\Sigma^+$ baryons produced from polarized $u$ quarks to be polarized. As we will review, more sophisticated approaches also predict the $\Sigma$ baryons to be much more sensitive to the polarizations of the $u$ and/or $d$ quarks than the $\Lambda$ baryons.

It is also important to note that the experimental question of $\Sigma^+$ reconstruction at the LHC that we address in this work will be relevant to studies of fragmentation functions involving the $\Sigma^+$ at the high-luminosity phase of the LHC (continuing and complementing the LEP program~\cite{Abreu:1995qx,Alexander:1996qi,Acciarri:2000zf}), even outside the context of the polarization. In fact, certain measurements of $b$-quark fragmentation in $t\bar t$ events have already begun in ATLAS~\cite{ATLAS:2022miz} and CMS~\cite{CMS-PAS-TOP-18-012}.

Finally, the $\Sigma^+$ decay that we will study here has an interesting signature, a kinked track, which will require a dedicated reconstruction procedure that is not currently being employed in any ATLAS or CMS analyses. Kinked tracks may appear also due to particles beyond the Standard Model (e.g., in various supersymmetric scenarios~\cite{Barr:2002ex,Asai:2011wy,Jung:2015boa}), and the $\Sigma^+$ analyses we propose here could lay the ground for future searches for such new physics scenarios.

\section{Theoretical status}

The probabilities for a given quark to produce a given hadron with a specific momentum fraction $z$ are described by fragmentation functions. Polarized fragmentation functions take the spins of the quark and hadron into account as well (see, e.g.,~\cite{deFlorian:1997zj}). Analogous to the parton distribution functions (PDFs), the fragmentation functions are not computable analytically because they depend on non-perturbative QCD physics. However, they can be measured in some process and then, again like the PDFs, used for describing another process after they are evolved to the appropriate scale. The evolution equations are known from perturbative QCD at the next-to-leading order (see, e.g,~\cite{Stratmann:1996hn}). The LEP experiments measured the fragmentation of quarks from $Z$ decays into $\Sigma^+$~\cite{Abreu:1995qx,Alexander:1996qi,Acciarri:2000zf}, as well as $\Sigma^-$~\cite{Abreu:1995qx,Alexander:1996qi,Abreu:2000nu} and $\Sigma^0$~\cite{Adam:1996hw,Barate:1996fi,Alexander:1996qi,Acciarri:2000zf}, without measuring the polarization. Unfortunately, even the unpolarized fragmentation functions for specific quark flavors into the $\Sigma^+$ cannot be extracted from these results directly due to the flavor-inclusive nature of the measurements. Flavor-specific fragmentation functions were extracted~\cite{Bourrely:2003wi} from part of these data~\cite{Alexander:1996qi} by assuming that they behave according to the so-called statistical model. In view of the reliance on this assumption, one should not necessarily be too discouraged by the result of~\cite{Alexander:1996qi} that the probability for a $u$ quark to produce a $\Sigma^+$ is lower than its probability to produce a $\Lambda$ (in the relevant range of momentum fractions) by a factor of $\sim 10$. (For the $s$ quark, the reduction is by a factor of $\sim 2$.) In any case, as will become clear in the next section, $t\bar t$ samples at the LHC provide an opportunity to measure these numbers more directly.

Phenomenological expectations regarding the salient features of polarized fragmentation functions are usually based on the spin distributions in the hadrons (polarized PDFs). For the nucleons, the spin distributions are known experimentally~\cite{Aidala:2012mv,Adolph:2015saz}. For the $\Lambda$ and $\Sigma^+$ baryons that we discuss here, they can be estimated either by applying the SU(3) flavor symmetry (which is only approximate because $m_s/\Lambda_{\rm QCD}$ is non-negligible) to the nucleon data~\cite{Burkardt:1993zh,Jaffe:1996wp}, or from a phenomenological model of QCD, or by a lattice simulation~\cite{Gockeler:2002uh,Chambers:2014qaa}. While one does not expect a simple general relation between PDFs and fragmentation functions since even their renormalization group evolution is different, an approximate relation at a low scale can be argued to exist~\cite{Barone:2000tx}.

The spin distributions of quarks in a fully-polarized spin-1/2 hadron $h$ are described at the simplest level in terms of the quantities
\beq
\Delta q_h = \int dx\, \left[q^\uparrow_h(x) - q^\downarrow_h(x)\right] ,
\eeq
where $q^s_h(x)$ describes the probability to find a quark of flavor $q$ and momentum fraction $x$ in the hadron $h$, with polarization projection $s$ on the direction of the hadron spin. If the hadron spin were determined by the valence quark spins alone, one would have $\sum_q \Delta q_h = 1$. This is the case in the na\"{i}ve quark model, where
\beq
\Delta u_p = \frac43 \,,\quad
\Delta d_p = -\frac13 \,,
\eeq
\beq
\Delta u_\Lambda = \Delta d_\Lambda = 0 \,,\quad
\Delta s_\Lambda = 1 \,,
\eeq
\beq
\Delta u_{\Sigma^+} = \frac43 \,,\qquad
\Delta s_{\Sigma^+} = -\frac13 \,.
\eeq
In reality, $\sum_q \Delta q_h \neq 1$ because orbital angular momentum, gluons, and sea quarks and antiquarks also contribute to the hadron spin. For the proton, the spin distributions are measured to be~\cite{Adolph:2015saz}
\beq
\Delta U_p = 0.835 \pm 0.015 \,,\qquad
\Delta D_p = -0.435 \pm 0.015 \,,\qquad
\Delta S_p = -0.095 \pm 0.015 \,,
\label{DIS-results}
\eeq
where the usage of capital $U$, $D$, $S$ symbols indicates that antiquark contributions are included. The SU(3) flavor symmetry relations~\cite{Burkardt:1993zh,Jaffe:1996wp,Boros:1998kc,Ashery:1999am}
\beq
\Delta U_\Lambda = \Delta D_\Lambda = \frac16\Delta U_p + \frac23\Delta D_p + \frac16\Delta S_p \,,
\eeq
\beq
\Delta S_\Lambda = \frac23\Delta U_p - \frac13\Delta D_p + \frac23\Delta S_p
\eeq
and
\beq
\Delta U_{\Sigma^+} = \Delta U_p\,,\qquad
\Delta D_{\Sigma^+} = \Delta S_p\,,\qquad
\Delta S_{\Sigma^+} = \Delta D_p
\eeq
then predict
\beq
\Delta U_\Lambda = \Delta D_\Lambda \approx -0.17 \,,\qquad
\Delta S_\Lambda \approx 0.64 \,,
\eeq
\beq
\Delta U_{\Sigma^+} \approx 0.835 \,,\qquad
\Delta D_{\Sigma^+} \approx -0.095 \,,\qquad
\Delta S_{\Sigma^+} \approx -0.435 \,.
\eeq
Lattice QCD simulations of the proton~\cite{Alexandrou:2020sml} give
\beq
\Delta U_p =  0.864 \pm 0.016 \,,\qquad
\Delta D_p = -0.426 \pm 0.016 \,,\qquad
\Delta S_p = -0.046 \pm 0.008 \,,
\eeq
in a reasonable agreement with the measured values in eq.~\eqref{DIS-results}. For the $\Lambda$ and $\Sigma^+$, available lattice results~\cite{Gockeler:2002uh,Chambers:2014qaa} include only connected contributions (i.e.\ valence quarks) and they are
\beq
\Delta u_\Lambda = \Delta d_\Lambda \approx -0.02 \,,\qquad
\Delta s_\Lambda \approx 0.68 \,,
\eeq
\beq
\Delta u_{\Sigma^+} \approx 0.81 \,,\qquad
\Delta s_{\Sigma^+} \approx -0.25 \,.
\eeq
We see that regardless of the approach taken, the spin distributions differ significantly between the $\Lambda$ and the $\Sigma^+$. The spin of the $\Lambda$ is predominantly on the $s$ quark, while the spin of the $\Sigma^+$ is predominantly on the $u$ quarks. It is reasonable to expect analogous differences in the polarization transfer in fragmentation. In particular, polarization transfer from the $u$ quark to the $\Sigma^+$ is expected to be more significant than to the $\Lambda$. Note that to estimate the polarization transfer factor for $u \to \Sigma^+$, one should divide $\Delta u_{\Sigma^+}$ by 2 because there are two valence $u$ quarks in the $\Sigma^+$. However, even after accounting for that, the $u$-quark polarization transfer to the $\Sigma^+$ is expected to be better than in the $\Lambda$ case by more than a factor of 2, or perhaps even a bigger factor if a significant fraction of $\Delta U_\Lambda$ is due to antiquarks, which is a plausible scenario~\cite{Jaffe:1996wp}.\footnote{For various approaches to estimating the antiquark contributions, see e.g.~\cite{Gluck:2000dy,deFlorian:2009vb,Nocera:2014gqa,Chen:2016utp}.} The $s$-quark polarization transfer to the $\Sigma^+$ is somewhat worse than in the $\Lambda$ case, but still sizable and includes a sign flip.

More generally, both the quark spin distributions in hadrons and the quark-to-hadron polarization transfer in fragmentation depend on the momentum fraction of the quark in the hadron ($x$) or the quark momentum fraction taken by the hadron ($z$), respectively.\footnote{For attempts to predict the $z$ dependence of the various fragmentation functions based on phenomenological models, see~\cite{Ma:1999wp,Ma:2000uu,Ma:2000cg,Ma:2001ri,Yang:2001yda,Yang:2001sy,Liu:2011ha,Chi:2013hka}.} The polarization transfer from the $u$ and $d$ quarks to the $\Lambda$ might have a particularly strong $z$ dependence since the inclusive numbers, as quoted above, are small.\footnote{Since the contributions are bounded between $-1$ and $1$, an ${\cal O}(1)$ inclusive number would suggest that the relative variation of the contributions with $x$ is not very large. On the other hand, a small inclusive number does not preclude significant $x$ dependence. In fact, significant $x$ dependence of the $u$-quark polarization in the $\Lambda$, including even a change of sign, is predicted by several phenomenological models (see, e.g., figure~2 in ref.~\cite{Liu:2011ha}).} For the polarization transfer from the $u$ to the $\Sigma^+$, it is likely much less of an issue.

Let us now compile all the theoretical factors to compare the prospects of the $\Sigma^+$-based and $\Lambda$-based measurements for identical samples of $s$ and $u$ jets. The statistical significance of the results will behave as
$$
n_\sigma(q,h) \propto |\alpha_h\,\Delta\hat q_h|\sqrt{N(q,h)} \;,
$$
where $q = s$ or $u$, $h = \Lambda$ or $\Sigma^+$, $\alpha_h$ is the spin analyzing power of the decay, $\Delta\hat q_h$ is the polarization transfer factor that we estimate from the spin distributions $\Delta q_h$ discussed above, and $N(q,h)$ is the number of decays available for the analysis, which is proportional to the $q \to h$ fragmentation fraction and the $h$ decay branching fraction. Collecting the numbers quoted above, we obtain
$$
\frac{n_\sigma(s,\Sigma^+)}{n_\sigma(s,\Lambda)} \sim 0.4 \,,\qquad
\frac{n_\sigma(u,\Sigma^+)}{n_\sigma(u,\Lambda)} \sim 1 \,.
$$
We see that the prospects of $\Sigma^+$-based analyses (at the theoretical level, before accounting for backgrounds and experimental factors) are comparable to those of $\Lambda$-based analyses, particularly for $u$ quarks. It is therefore worth exploring the experimental feasibility of both approaches, especially in view of the fact that they will provide complementary information.

\section{Experimental opportunities}

So far, neither ATLAS nor CMS have reported any analyses involving a reconstruction of the $\Sigma^+ \to p \pi^0$ decays. We hope this work will provide a motivation for such analyses.

Since the $\Sigma^+$ is long-lived ($c\tau \approx 2.4\,$cm), its signature would often include a kink in the tracker (formed by the $\Sigma^+$ and $p$ tracks) as well as a $\pi^0$ candidate in the electromagnetic calorimeter. Once the decay kinematics is reconstructed, the $\Sigma^+$ polarization, $\pol(\Sigma^+)$, can be obtained using the fact that the angular distribution of the proton momentum in the $\Sigma^+$ rest frame behaves as
\beq
\frac{1}{\Gamma}\,\frac{d\,\Gamma}{d \cos \vartheta} = \frac12\left(1 + \alpha\,\pol(\Sigma^+)\cos\vartheta\right) ,
\label{spin-analyzing}
\eeq
where $\vartheta$ is the proton momentum angle relative to the polarization axis and $\alpha = -0.982 \pm 0.014$~\cite{PDG}. For longitudinal polarization, this axis is the $\Sigma^+$ direction of motion in the lab frame.

\subsection{Kinked track}

We are mostly interested in $\Sigma^+$ baryons carrying about $30$--$50\%$ of the jet's momentum. Much softer $\Sigma^+$ baryons are common in secondary fragmentation processes, where they will usually be unrelated to the original quark, while harder $\Sigma^+$ baryons are very rare. This means that a typical $p_T$ of the $\Sigma^+$, for example in jets from $W$ decays in $t\bar t$ samples (which we will consider in the following), will be $p_T \sim 15$~GeV. Then the kink will occur at a typical distance of
\beq
r = c\tau\,\frac{p_T}{m} \sim 30~\mbox{cm}
\eeq
from the beam axis. This falls near the beginning of the SCT (Semiconductor Tracker) in ATLAS or the TIB (Tracker Inner Barrel) in CMS. A typical $\Sigma^+$ track will therefore pass through all the layers of the Pixel Detector in either ATLAS or CMS, and often also the first layer(s) of the SCT or TIB.\footnote{The Pixel Detector layers in ATLAS are positioned at $r = 3.3,\, 5.05,\, 8.85,\, 12.25\,\mbox{cm}$~\cite{Pernegger:2015qqa}, and in CMS at $r = 3.0,\, 6.8,\, 10.2,\, 16.0\,\mbox{cm}$~\cite{CMS:2012sda}.
The rest of the barrel tracker in ATLAS consists of the Semiconductor Tracker (SCT) between $r = 30$ and $52$~cm and the Transition Radiation Tracker (TRT) between $r = 56$ and $107$~cm~\cite{ATLAS-TDR-4}. The rest of the barrel tracker in CMS consists of the strip tracker inner barrel (TIB) between $r = 20$ and $55$~cm and the strip tracker outer barrel (TOB) between $r = 55$ and $116$~cm~\cite{Chatrchyan:2014fea}.} The proton track will pass through the rest of the tracker.

While there have been no analyses targeting kinks so far (even though various motivations exist~\cite{Barr:2002ex,Asai:2011wy,Jung:2015boa} in addition to ours), searches for \emph{disappearing tracks} are being conducted by ATLAS~\cite{Aaboud:2017mpt,ATLAS:2022rme} and CMS~\cite{Sirunyan:2018ldc,CMS:2019ybf,CMS:2020atg}. The Run~2 ATLAS searches~\cite{Aaboud:2017mpt,ATLAS:2022rme} reconstruct disappearing tracks with Pixel Detector hits alone, similar to what would be required for reconstructing the $\Sigma^+$ track. The track $p_T$ threshold in~\cite{Aaboud:2017mpt,ATLAS:2022rme} is $20$~GeV, comparable to the $p_T$ of the $\Sigma^+$ in the samples of interest. In the Run~2 CMS searches~\cite{CMS:2019ybf,CMS:2020atg}, multiple options for the disappearing track length were addressed, including tracks that extend beyond the pixel detector, with $p_T$ values down to $15$~GeV~\cite{CMS:2019ybf}. For muon tracks from $Z$ decays, with typical $p_T \sim 45$~GeV, the pixel-only tracks in ATLAS have $p_T$ resolution of about $60\%$~\cite{Aaboud:2017mpt}, so for a $p_T \sim 15$~GeV pixel-only track we should expect $\sim 20\%$ resolution in $p_T$. Similar resolution is expected in CMS, where the pixel layers have similar geometry, and the bigger pixel size than in ATLAS is compensated by a larger magnetic field.\footnote{The pixel size in the ATLAS Pixel Detector is $50 \times 250\,\mu\mbox{m}^2$ for the innermost layer and $50 \times 400\,\mu\mbox{m}^2$ for the other layers~\cite{Pernegger:2015qqa}, and in CMS it is $100 \times 150\,\mu\mbox{m}^2$~\cite{CMS:2012sda}, where the first dimension is the transverse one, which is the one relevant for the $p_T$ measurement. In ATLAS's SCT and in the first layers of CMS's TIB the pitch is $80\,\mu\mbox{m}$~\cite{ATLAS-TDR-4,Chatrchyan:2014fea}. The magnetic field is $2$~T in ATLAS and $3.8$~T in CMS.} Significantly better resolution, of $\sim 5\%$, will be obtained for tracks reaching the first layers of the SCT or TIB, which will be common in our case. Further improvement will be attained in the high-luminosity phase of the LHC, once the upgrades to $25 \times 100\,\mu\mbox{m}^2$ pixels in ATLAS~\cite{ATLAS-TDR-030} and $50 \times 50\,\mu\mbox{m}^2$ or $25 \times 100\,\mu\mbox{m}^2$ pixels in CMS~\cite{CMS-TDR-15-02} are implemented.

The $\Sigma^+$ kink angle $\varphi$ is given by
\beq
\sin\varphi = \frac{1}{\displaystyle\sqrt{1 + \frac{\gamma^2\left(\cos\vartheta + \beta E_p/p_p\right)^2}{\sin^2\vartheta}}}
\simeq \frac{\sin\vartheta}{\gamma\left(\cos\vartheta + E_p/p_p\right)} \;,
\label{kink-angle}
\eeq
where $\vartheta$ is the rest-frame decay angle from eq.~\eqref{spin-analyzing}, $\beta$ is the $\Sigma^+$ velocity, $\gamma = 1/\sqrt{1-\beta^2}\,$ is the corresponding boost factor, and $E_p$ and $p_p$ are the energy and momentum of the proton in the $\Sigma^+$ rest frame, where one has $E_p / p_p \approx 5.1$ (see appendix~\ref{app-kinematics} for these and other details of the kinematics). In the last step in eq.~\eqref{kink-angle}, we took the ultra-relativistic limit $\gamma \gg 1$. We see that the typical kink angle in our case ($\gamma \sim 15$) will be $\varphi \sim 10$\,mrad (see also table~\ref{tab:kinky-decays}). It is measurable.\footnote{For example, even the Run-1 CMS three-layer Pixel Detector alone was capable of measuring the track momentum parameters $\phi$ and $\theta$ with resolutions of roughly $3$\,mrad and $1$\,mrad, respectively, for tracks with $p_T \sim 10$~GeV~\cite{Chatrchyan:2014fea}.} However, since $E_p/p_p \approx 5.1 \gg 1$, eq.~\eqref{kink-angle} implies that the kink angle $\varphi$ is not very sensitive to the forward/backwardness of the proton (i.e., the sign of $\cos\vartheta$). In fact, most values of $\varphi$ can be obtained from two very different values of $\cos\vartheta$ with opposite signs. This happens because the proton is nonrelativistic in the $\Sigma^+$ rest frame. As a result, the longitudinal component (with respect to the $\Sigma^+$ direction of motion) of its momentum in the lab frame is dominated by the boost of its mass (rather than its 3-momentum) from the $\Sigma^+$ rest frame. Instead of using $\varphi$ itself, it may be convenient to extract $\cos\vartheta$ via
\beq
\cos\vartheta = \frac{\beta}{p_p}\left(\frac{p_p'}{p_{\Sigma^+}'}m_{\Sigma^+}\cos\varphi - E_p\right)
\approx \beta\left(6.3\,\frac{p_p'}{p_{\Sigma^+}'}\cos\varphi - 5.1\right)
\simeq 6.3\,\frac{p_p'}{p_{\Sigma^+}'} - 5.1 \,,
\eeq
where the primed ($'$) quantities denote lab-frame momenta, and in the last step we assumed the $\Sigma^+$ to be relativistic. Since the range of $\cos\vartheta$ is $[-1,1]$, the momenta of the proton and $\Sigma^+$ need to be measured with precision of $\sim 10\%$ in order for them to provide useful information about $\cos\vartheta$. Based on the discussion in the previous paragraph, it seems realistic despite the shortness of the $\Sigma^+$ track.

\begin{table}[t]
\begin{center}
\begin{tabular}{c|c|c|c}
Decay                                      & \multicolumn{3}{|c}{$\displaystyle\left(\frac{P}{15\,\mbox{GeV}}\right)\times\varphi$ [mrad]} \\\cline{2-4}
\qquad\qquad$\vartheta =$                  & $\quad\pi/4\quad$ & $\quad\pi/2\quad$ & $\quad3\pi/4\quad$ \\\hline\hline
$\Sigma^+ \to p \pi^0$                     & 10 & 16 & 13 \\\hline
$\Sigma^+ \to n\pi^+$                      & 29 & 63 & 103 \\
$\pi^+ \to \mu^+\nu_\mu$                   &  1 & 3  &  2 \\
$K^+ \to \mu^+\nu_\mu$                     & 13 & 30 & 60 \\
$K^+ \to \pi^+\pi^0$                       & 12 & 27 & 46 \\
$\overline\Sigma^+ \to \overline n\pi^+$   & 29 & 65 & 107 \\
$\overline\Xi^+ \to \pi^+\overline\Lambda$ & 29 & 62 & 88 \\
\end{tabular}
\end{center}
\caption{The kink angle $\varphi$ for three values of the rest-frame decay angle $\vartheta = \pi/4$, $\pi/2$, $3\pi/4$, where $P$ is the lab-frame momentum of the mother hadron, which is assumed to be relativistic, $\gamma \gg 1$.}
\label{tab:kinky-decays}
\end{table}

There will be background due to kinks produced by other decays of charged hadrons: $\Sigma^+ \to n\pi^+$, $\pi^+ \to \mu^+\nu_\mu$, $K^+ \to \mu^+\nu_\mu$, $K^+ \to \pi^+\pi^0$, $\overline\Sigma^+ \to \overline n\pi^+$ and $\,\overline\Xi^+ \to \pi^+\overline\Lambda$. We note that among these, only $\Sigma^+ \to n\pi^+$ can arise from a primary strange quark. Nonprimary hadrons (namely ones not carrying the original quark) will typically be soft relative to the jet, thus not of interest to us in any case. Therefore, for a high-purity sample of strange quarks, like the one obtained for instance in the $t\bar t$ selection described below, $\Sigma^+ \to n\pi^+$ will make the dominant contribution to this background. Luckily, this contribution can be largely eliminated using the fact that the kink angle in this case is usually much larger than in the decay of interest, as shown in table~\ref{tab:kinky-decays}. For a sample of up quarks, the typical size of the kink angle can also be used to reduce some of the contributions.

Kinks can also be created by interactions of charged hadrons with the detector material---either an elastic scattering with a large momentum transfer or an inelastic scattering such as $\pi^+ + n \to \pi^0 + p$. However, in the elastic scattering case the scattering angle is typically much smaller than the $\Sigma^+$ kink angle. The width of the scattering angle distribution is given by (section~34.3 of~\cite{PDG})
\beq
\varphi_{\rm rms} \simeq \frac{13.6~\mbox{MeV}}{P}\sqrt{2\frac{x}{X_0}} \;,
\eeq
where $P$ is the hadron momentum and $x/X_0$ is the medium thickness in radiation lengths. For ATLAS,\footnote{We could not find an analogous number for CMS, but it seems~\cite{CMS:2012sda,CMS:2018wqs,Chatrchyan:2014fea} that there is no reason to expect a significantly bigger number.} in the range $15\,{\rm cm} < r < 35\,{\rm cm}$, we have $x/X_0 \sim 0.05$~\cite{Aaboud:2017pjd}, thus
\beq
\varphi_{\rm rms} \sim \left(\frac{15\,\mbox{GeV}}{P}\right) \times 0.3\,\mbox{mrad} \,,
\eeq
which is more than an order of magnitude smaller than the kink angle in the $\Sigma^+ \to p\pi^0$ decay (see table~\ref{tab:kinky-decays}).

Part of the inelastic scattering background can be avoided by requiring the momentum of the second part of the track to be consistent with the proton produced in the decay, which is expected to carry between $65\%$--$96\%$ of the $\Sigma^+$ momentum. Both of the material-induced backgrounds can be reduced significantly, without too much efficiency loss to the signal, by vetoing kinks occurring near detector layers. A detailed simulation of the ATLAS and/or CMS detector, which is not available to the author, would be needed to define the details of this procedure and estimate the background that will remain.

\subsection{$\pi^0$}

It is not straightforward to associate a $\pi^0$ with a candidate $\Sigma^+$ decay because the trackless $\pi^0$ cannot be assigned to a specific vertex, and there are various ways in which unrelated $\pi^0$s can be produced inside a jet or from pile-up, especially considering that our $\pi^0$ is very soft ($E_{\pi^0} \sim 3$~GeV for $p_T^{\Sigma^+} \sim 15$~GeV). Note though that if the $\Sigma^+$ and $p$ momenta are well-measured, measuring the $\pi^0$ is not essential for reconstructing the decay. Still, it may be useful to include a loose requirement that a $\pi^0$ candidate with the expected energy\footnote{The electromagnetic calorimeter resolution is given by
$$\frac{\sigma_E}{E} = \frac{a}{\sqrt{E/\mbox{GeV}}} \oplus \frac{b}{E/\mbox{GeV}} \oplus c\,,$$
where $a = 10.1 \pm 0.1\%$, $b \approx 8\%$, $c = 0.17 \pm 0.04\%$ in ATLAS~\cite{Aharrouche:2006nf,Aad:2008zzm,Aad:2014nim}, and $a \approx 2.8\%$, $b \approx 12\%$, $c \approx 0.3\%$ in CMS~\cite{Chatrchyan:2013dga}, so the energy of the $\pi^0$ can be measured with a precision of around $5\%$.} and position in the calorimeter be present in the jet, to suppress backgrounds. For $p_T^{\Sigma^+} \sim 15$~GeV, a typical deviation of the $\pi^0$ from the $\Sigma^+$ direction will be $80$~mrad.

It is also worth checking whether one can identify the non-pointing and/or delayed nature of the $\pi^0$ photons related to the large displacement of the $\Sigma^+$ decay vertex, along the lines of~\cite{Aad:2014gfa,CMS:2019zxa,ATLAS:2022vhr,CMS-PAS-EXO-14-017}, which would eliminate much of the background $\pi^0$s. These analyses determine the flight direction of the photon based on its shower shape in the electromagnetic calorimeter~\cite{Aad:2014gfa,CMS:2019zxa,ATLAS:2022vhr} (or $e^+e^-$ tracks in the case of converted photons~\cite{CMS-PAS-EXO-14-017}) and/or measure the delay in the arrival time relative to a photon that would arrive directly from the primary vertex~\cite{Aad:2014gfa,CMS:2019zxa,ATLAS:2022vhr}.

In principle, it might also be possible to reconstruct the $\Sigma^+$ decay using just the proton and the $\pi^0$, without relying on a direct measurement of the $\Sigma^+$ momentum. This was done (although with relatively low efficiencies) by OPAL~\cite{Alexander:1996qi} and L3~\cite{Acciarri:2000zf} in samples of hadronic $Z$ decays at LEP and by LHCb (as a control channel for another measurement) in inclusive QCD samples~\cite{Aaij:2017ddf}. However, given the potential difficulty in identifying the relevant soft $\pi^0$, we leave the exploration of this interesting alternative beyond the scope of this paper.

\subsection{Proposed analyses in $t\bar t$ events}

We propose to measure the polarization transfer from the $s$ and $u$ quarks to the $\Sigma^+$ baryon using $t\bar t$ samples in ATLAS and CMS.

We envision starting with a standard $t\bar t$ event selection in the lepton+jets channel. Such a selection usually requires the event to contain one isolated electron or muon, at least four jets, one or two of which are required to be $b$ tagged, and satisfy a mild cut on the missing energy and/or the transverse mass of the leptonically decaying $W$. After a kinematic reconstruction of the event, information about the flavor of the jets from the hadronic $W$ decay can be obtained using charm tagging. Demanding a charm tag will produce high-purity and high-efficiency samples of $c$ and $s$ jets from the
\beq
W^+ \to c\bar s
\eeq
decays. For example, charm tagging working points with $40\%$ (or $30\%$) efficiency for $c$ jets and $\sim 3\%$ (or $1\%$) efficiency for light jets are possible in both ATLAS~\cite{ATLAS:2022qxm} and CMS~\cite{CMS:2021scf}. Events with $\Sigma^+$ baryons in the $s$-jet candidates will then allow measuring the
\beq
s \to \Sigma^+
\eeq
polarization transfer.

Alternatively, vetoing on a loose charm tag will provide high-efficiency samples of
\beq
W^+ \to u\bar d
\eeq
decays, which can be used for measuring the
\beq
u \to \Sigma^+
\eeq
polarization transfer. The sample will, however, contain a significant contamination from $W^+ \to c\bar s$. For example, with the charm tagging algorithm of ref.~\cite{CMS:2021scf}, only $80\%$ of the $c\bar s$ decays will be successfully vetoed while keeping $50\%$ of the $u\bar d$ decays. The impact of the contamination from $c\bar s$ will be further enhanced by $\Sigma^+$ baryons being more readily produced from $c$ quarks than from $u$ quarks, on which we will elaborate below. Production from the $\bar d$ or $\bar s$ antiquarks is not an issue: their jets will not contribute many energetic $\Sigma^+$ baryons because the $\Sigma^+$ does not contain valence antiquarks. Here we assume that the $W$ charge, and correspondingly the charge expected for the $\Sigma^+$, is determined by the lepton on the other side of the event.

Let us now estimate the statistics available for these analyses. A standard $t\bar t$ selection in the lepton+jets channel in the ATLAS Run~2 dataset~\cite{ATLAS:2020aln} results in $5.6 \times 10^6$ $t\bar t$ events and a background of $1.1 \times 10^6$ events from other processes. The CMS analysis in the same final state~\cite{CMS:2021vhb} also describes how the event kinematics can be reconstructed to determine, in particular, which of the observed jets come from the $W$ decay. We envision a similar procedure for our proposed analysis. Charm tagging, which we would use anyway to select the sample of strange jets that accompany the charm jets in the $W^+ \to c\bar s$ decays, will aid the reconstruction. Since approximately half of the hadronic $W$ decays produce $c\bar s$, and assuming $40\%$ charm tagging efficiency, we obtain a sample of about $1.1 \times 10^6$ strange jets. Based on the fragmentation functions from ref.~\cite{Bourrely:2003wi}, the probability for a strange quark to produce an energetic $\Sigma^+$, which we define as carrying at least $30\%$ of the jet momentum (i.e.,~$z > 0.3$), is about $1.6\%$. Taking into account also the branching fraction for $\Sigma^+ \to p \pi^0$, we are left with about $9000$ such decays in the Run~2 samples. Even if only a fraction of the kinks are reconstructed and certain additional mild cuts are applied to reduce backgrounds, there will likely still remain a sizable $s \to \Sigma^+$ sample to study even in the data that has already been collected.

For $u \to \Sigma^+$, the statistics is more challenging because the corresponding fragmentation fraction for $z > 0.3$ is only about $2.2 \times 10^{-4}$~\cite{Bourrely:2003wi}. Estimating as in the previous case and assuming $50\%$ efficiency for passing the loose charm tag veto, we are left with about $160$ decays in the Run~2 dataset before accounting for the kink reconstruction efficiency and the need to deal with significant backgrounds in this case. As mentioned above, there is a sizable contamination due to $c$ jets that evade the charm tag veto. They contribute primarily via $\Lambda_c^+ \to \Sigma^+ + X$. The fragmentation fraction for $c \to \Lambda_c^+$ is about $6\%$ (at high $p_T$)~\cite{Gladilin:2014tba,Lisovyi:2015uqa,Kniehl:2020szu} and the branching fraction of $\Lambda_c^+ \to \Sigma^+\,$(+~anything) is about $11\%$~\cite{PDG}.\footnote{Known contributions to BR$(\Lambda_c^+ \to \Sigma^+{\rm anything})$ are
BR$(\Lambda_c^+ \to \Sigma^+\pi^+\pi^-) = (4.50 \pm 0.25)\%$, 
BR$(\Lambda_c^+ \to \Sigma^+\omega) = (1.70 \pm 0.21)\%$,
BR$(\Lambda_c^+ \to \Sigma^+\pi^0\pi^0) = (1.55 \pm 0.15)\%$,
BR$(\Lambda_c^+ \to \Sigma^+\eta') = (1.5 \pm 0.6)\%$,
BR$(\Lambda_c^+ \to \Sigma^+\pi^0) = (1.25 \pm 0.10)\%$,
BR$(\Lambda_c^+ \to \Sigma^+\eta) = (0.44 \pm 0.20)\%$,
BR$(\Lambda_c^+ \to \Sigma^+K^+K^-) = (0.35 \pm 0.04)\%$,
BR$(\Lambda_c^+ \to \Sigma^+K^+\pi^-) = (0.21 \pm 0.06)\%$~\cite{PDG}. They add up to $(11.5 \pm 0.7)\%$. It is also known that BR$(\Lambda_c^+ \to \Sigma^\pm\,{\rm anything}) = (10\pm5)\%$~\cite{ParticleDataGroup:2018ovx}.} About half of these decays (such as the most common one, $\Lambda_c^+ \to \Sigma^+\pi^+\pi^-$) produce only charged particles in addition to the $\Sigma^+$ and can therefore be vetoed via $\Lambda_c^+$ mass reconstruction. This leaves us with a probability of $3 \times 10^{-3}$ for a $c$ jet to contribute a $\Sigma^+$ while evading the charm tag veto and the $\Lambda_c^+$ reconstruction veto. A significant fraction of these $\Sigma^+$ baryons are likely sufficiently energetic to compete with the $u \to \Sigma^+$ signal which has a fragmentation fraction of $2.2 \times 10^{-4}$ for $z > 0.3$. Since the number of $c$ jets entering the sample after the charm tag veto is smaller than the number of $u$ jets by only a factor of about $2.5$, the $c$-jet background is nonnegligible, but at the same time it is not prohibitively large. One may consider using the deviation of the $\Sigma^+$ trajectory from the jet axis as a handle to further reduce such contributions. While the statistics of $160$ signal events (before accounting for the reconstruction efficiency and addressing backgrounds) is problematic, it will increase by a factor of at least $20$ at the high-luminosity LHC, along with detector upgrades, and then a meaningful measurement may become possible for this process as well.

\section{Summary}

Top-quark pair production ($pp \to t\bar t$) samples at the LHC are a high-statistics and relatively clean source of polarized quarks of various flavors --- $b$, $c$, $s$, $d$, $u$ --- that can be studied by ATLAS and CMS. Decay angular distributions of energetic baryons inside the jets are sensitive to the original quark polarization. Previous proposals focused on the $\Lambda_b$ and $\Lambda_c$ baryons from the $b$ and $c$ quarks, respectively~\cite{Galanti:2015pqa}, and the $\Lambda$ baryons from the $s$ quarks~\cite{Kats:2015cna}. In the current paper we considered the production of $\Sigma^+$ baryons from the $s$ or $u$ quarks, which can provide complementary information.

We proposed to use the kinked track signature of the $\Sigma^+ \to p \pi^0$ decay (possibly along with the electromagnetic signature of the displaced $\pi^0$) to identify the $\Sigma^+$ baryons in $s$ and $u$ jet candidates and measure their polarization. Measuring the $s \to \Sigma^+$ polarization transfer seems feasible with existing data, while the $u \to \Sigma^+$ measurement, where statistics is lower and backgrounds are larger, will probably become possible only at the high-luminosity LHC.

Apart from measuring polarization transfer, analyses with kink reconstruction could be useful for measuring fragmentation functions involving the $\Sigma^+$ and other hadrons with kinked track signatures, as well as searching for certain scenarios in which particles beyond the Standard Model produce kinked tracks~\cite{Barr:2002ex,Asai:2011wy,Jung:2015boa}. We hope this paper will further motivate such analyses.

\acknowledgments
The author would like to thank Abner Soffer, whose important comments and insights contributed to this work. This research was supported in part by the Israel Science Foundation (grants no.~780/17 and 1666/22) and the United States - Israel Binational Science Foundation (grant no.~2018257).

\appendix

\section{Kinematics}
\label{app-kinematics}

This appendix collects formulas and numbers relevant to the kinematics of the $\Sigma^+ \to p \pi^0$ decay. The symbols $\Sigma$ and $\pi$ below refer to the $\Sigma^+$ and $\pi^0$, respectively. Energies and momenta in the lab frame will be denoted with a prime ($'$), so transformations from the $\Sigma$ rest frame are written as
\beq
E' = \gamma \left(E + \beta p_\|\right) \,,\qquad
p_\|' = \gamma \left(p_\| + \beta E\right) \,,\qquad
\vec p_\perp' = \vec p_\perp \,,
\eeq
where $\beta$ is the $\Sigma$ velocity in the lab frame, $\gamma = 1/\sqrt{1-\beta^2}$ is the corresponding Lorentz factor, $\|$ and $\perp$ denote the momentum components with respect to the boost direction.

\subsection*{Quantities in the $\Sigma$ rest frame}

The energies and momenta are given by
\beq
E_\pi = \frac{m_\Sigma^2 + m_\pi^2 - m_p^2}{2m_\Sigma} \approx 0.232\,\mbox{GeV} \,,
\qquad
E_p = \frac{m_\Sigma^2 + m_p^2 - m_\pi^2}{2m_\Sigma} \approx 0.957\,\mbox{GeV} \,,
\eeq
\beq
p_p = p_\pi = \sqrt{E_\pi^2 - m_\pi^2} \approx 0.189\,\mbox{GeV} \,,
\eeq
where we used the masses~\cite{PDG}
\beq
m_\Sigma \approx 1.1894~\mbox{GeV} \,,\qquad
m_\pi \approx 0.1350~\mbox{GeV} \,,\qquad
m_p \approx 0.9383~\mbox{GeV} \,.
\eeq

\subsection*{Proton momentum and pion energy in the lab frame}

For a relativistic $\Sigma$ (i.e.\ $\beta \simeq 1$), the proton, which is produced non-relativistic in the $\Sigma$ rest frame ($v_p = p_p/E_p \approx 0.20$), acquires momentum of roughly
\beq
p_p' \simeq \gamma\beta E_p
= \frac{p_\Sigma'}{m_\Sigma}E_p \approx 0.80\, p_\Sigma'
\approx 12\,\mbox{GeV}\times\left(\frac{p_\Sigma'}{15\,\mbox{GeV}}\right)
\eeq
in the lab frame.

The pion energy in the lab frame is
\beq
E'_\pi = \gamma \left(E_\pi + \beta p_{\pi,\|}\right)
\simeq \frac{p'_\Sigma}{m_\Sigma} \left(E_\pi + p_{\pi,\|}\right) .
\eeq
Since $p_{\pi,\|}$ can be both positive and negative and its maximal magnitude is somewhat smaller than $E_\pi$, the typical value of the pion energy can be estimated to be
\beq
E'_\pi \sim \frac{p'_\Sigma}{m_\Sigma} E_\pi \approx 2.9\,\mbox{GeV}\times\left(\frac{p_\Sigma'}{15\,\mbox{GeV}}\right) .
\eeq

\subsection*{Kink angle as a function of the rest-frame decay angle}

Denoting the rest-frame decay angle (direction of proton momentum relative to the boost direction) by $\vartheta$, we have
\beq
p_{p,\|}' = \gamma \left(p_p \cos\vartheta + \beta E_p\right) \,,\qquad
p_{p,\perp}' = p_p \sin\vartheta \,,
\eeq
which gives the kink angle $\varphi$ as
\beq
\sin\varphi  = \frac{p_{p,\perp}'}{p_p'} = \frac{1}{\sqrt{1 + p_{p,\|}'^2/p_{p,\perp}'^2}} = \frac{1}{\displaystyle\sqrt{1 + \frac{\gamma^2\left(\cos\vartheta + \beta E_p/p_p\right)^2}{\sin^2\vartheta}}} \;,
\eeq
where $E_p / p_p \approx 5.1$, as we computed above. For $\gamma \gg 1$,
\beq
\sin\varphi \simeq \frac{\sin\vartheta}{\gamma\left(\cos\vartheta + E_p/p_p\right)}
\;\sim\; \frac{\sin\vartheta}{5\gamma} \,,
\eeq
where the last expression is a rough approximation.

A similar calculation applies to the pion. Since $E_\pi/p_\pi \approx 1.2$, a typical angle of the pion momentum relative to the $\Sigma^+$ momentum will be
\beq
\sin\varphi_\pi \sim \frac{1}{\gamma}
\simeq \frac{m_\Sigma}{p_\Sigma'}
\approx 0.08 \times \left(\frac{15\,\mbox{GeV}}{p_\Sigma'}\right) .
\eeq

\subsection*{Extraction of the decay angle from the measured momenta}

From the Lorentz transformation for $p_{p,\|}' = p_p'\cos\varphi$ we obtain
\beq
\cos\vartheta = \frac{1}{p_p}\left(\frac{p_p'\cos\varphi}{\gamma} - \beta E_p\right)
= \frac{\beta}{p_p}\left(\frac{p_p'}{p_\Sigma'}m_\Sigma\cos\varphi - E_p\right).
\eeq
Substituting the numerical values for the fixed quantities, we get
\beq
\cos\vartheta \approx \beta\left(6.3\,\frac{p_p'}{p_\Sigma'}\cos\varphi - 5.1\right)
\simeq 6.3\,\frac{p_p'}{p_\Sigma'} - 5.1 \,,
\eeq
where in the last step we assumed the $\Sigma$ to be relativistic (so that $\beta \simeq 1$ and $\cos\varphi \simeq 1$).

\renewcommand{\baselinestretch}{1.097}
\bibliographystyle{utphys}
\bibliography{SigmaPlus}

\end{document}